\documentclass[aps,twocolumn,showpacs,preprintnumbers,noshowpacs,superscriptaddress,6pt]{revtex4-2}

\usepackage[labelfont=bf]{caption}
\usepackage{euscript}
\usepackage{epsf}
\usepackage{epsfig}
\usepackage{subcaption}

\usepackage[utf8]{inputenc}
\usepackage{graphicx,epstopdf}
\usepackage{overpic}
\usepackage{textcomp}
\usepackage{multirow}
\usepackage{adjustbox}
\usepackage{natbib}

\usepackage{hyperref}

\usepackage{parskip}
\usepackage[T1]{fontenc}
\usepackage{dcolumn}

\usepackage{bm}
\usepackage{amsfonts}
\usepackage{amsmath}
\usepackage{amssymb}
\usepackage{siunitx}
\usepackage{booktabs}
\usepackage{mathtools}
\usepackage{flafter}
\usepackage{threeparttable}
\usepackage{subfiles}
\usepackage{import}
\usepackage{float}
\usepackage{makecell}
\usepackage{diagbox}
\usepackage{indentfirst}
\usepackage{mdframed}
\usepackage[strict]{chngpage}

\newmdenv{allfour}

\newmdenv[leftline=false,rightline=false]{topbot}

\newmdenv[topline=false,rightline=false]{leftbot}

\usepackage{ulem}

\usepackage{color, colortbl}

\definecolor{Gray}{gray}{0.9}
\definecolor{LightCyan}{rgb}{0.88,1,1}

\definecolor{BLUE}{rgb}{0.0,0.0,1.0}

\usepackage[T2A]{fontenc}

\graphicspath{{images/}}

\everymath{\displaystyle}
\renewcommand{\vec}[1]{{\mbox{\boldmath$#1$}}}

\usepackage{soul}

\begin{document}

\title{
Ionization potential and electron affinity of superheavy element 119: relativistic high-order coupled cluster study with QED corrections
}

\author{A.~R.~Saetgaraev}
\affiliation{Department of Physics, St. Petersburg State University, 7/9 Universitetskaya nab., 199034 St. Petersburg, Russia}

\author{L.~V.~Skripnikov}
\affiliation{Department of Physics, St. Petersburg State University, 7/9 Universitetskaya nab., 199034 St. Petersburg, Russia}
\affiliation{National Research Centre “Kurchatov Institute” B.P. Konstantinov Petersburg Nuclear Physics Institute, Gatchina,
Leningrad district 188300, Russia}

\author{I.~I.~Tupitsyn}
\email[]{i.tupitsyn@spbu.ru}
\affiliation{Department of Physics, St. Petersburg State University, 7/9 Universitetskaya nab., 199034 St. Petersburg, Russia}

\author{D.~P.~Usov}
\affiliation{Department of Physics, St. Petersburg State University, 7/9 Universitetskaya nab., 199034 St. Petersburg, Russia}

\author{A.~V.~Oleynichenko}
\affiliation{National Research Centre “Kurchatov Institute” B.P. Konstantinov Petersburg Nuclear Physics Institute, Gatchina,
Leningrad district 188300, Russia}
\affiliation{
Moscow Center for Advanced Studies, Kulakova Str. 20, Moscow, Russia}

\author{I.~M.~Savelyev}
\affiliation{Department of Physics, St. Petersburg State University, 7/9 Universitetskaya nab., 199034 St. Petersburg, Russia}

\author{N.~K.~Dulaev}
\affiliation{Department of Physics, St. Petersburg State University, 7/9 Universitetskaya nab., 199034 St. Petersburg, Russia}
\affiliation{National Research Centre “Kurchatov Institute” B.P. Konstantinov Petersburg Nuclear Physics Institute, Gatchina,
Leningrad district 188300, Russia}

\author{V.~M.~Shabaev}
\affiliation{Department of Physics, St. Petersburg State University, 7/9 Universitetskaya nab., 199034 St. Petersburg, Russia}
\affiliation{National Research Centre “Kurchatov Institute” B.P. Konstantinov Petersburg Nuclear Physics Institute, Gatchina,
Leningrad district 188300, Russia}

\date{\today}

\begin{abstract}
We report a highly accurate \textit{ab initio} study of the ionization potential (IP) and electron affinity (EA) of element 119. Electronic correlation effects are treated within the relativistic coupled cluster theory including excitations up to quadruples. The Gaunt electron–electron interaction and quantum electrodynamic (QED) corrections are taken into account. The role of high-order correlation effects is analyzed in detail. Our recommended values for the IP and EA are 4.7839(56) eV and 0.6750(71) eV, respectively. These results tighten previous estimates and provide a reference point for future experiments probing periodic-law trends beyond oganesson.

\end{abstract}

\maketitle
\section{Introduction}\label{sec:intro}

The superheavy elements (SHEs) with $\mathrm{Z = 104-118}$ were synthesized at JINR (Dubna), LBNL (Berkely), GSI (Darmstadt), and RIKEN (Japan)~\cite{Flerov1964, Zvara1969, Ghiorso1969, Ghiorso1970, FLEROV1971, Ghiorso1974, Oganessian1974, Hofmann2000, Morita2004, Oganessian1999, Stavsetra2009, Oganessian2000, Oganessian2010, Oganessian2013}. SHEs with $\mathrm{Z = 107-113}$ were synthesized \textit{via} cold fusion reactions, in which massive projectiles such as $\mathrm{^{58}Fe}$ or $\mathrm{^{64}Ni}$ were collided with $\mathrm{^{208}Pb}$ or $\mathrm{^{209}Bi}$ targets. However, this approach is no longer viable for producing elements with $\mathrm{Z \geq 114}$, as it fails to provide sufficient neutron excess in the evaporation residues and suffers from a significant decrease in cross section with increasing projectile mass~\cite{Oganessian2015}. Hence, SHEs with $\mathrm{Z = 114-118}$ were instead discovered through hot fusion reactions using $\mathrm{^{48}Ca}$-beam and various actinide targets. A target heavier than Cf (Z = 98) is difficult to produce, therefore for synthesis of SHEs with $\mathrm{Z \geq 119}$ different reaction should be utilized. Experiments on the synthesis of elements 119 (E119) and 120 (E120) were carried out at GSI~\cite{Hofmann2016, khuyagbaatar2020} and RIKEN \cite{Tanaka2022} using $\mathrm{^{50}Ti+^{249}Bk}$ and $\mathrm{^{51}V+^{248}Cm}$ nuclear reactions for E119 and $\mathrm{^{54}Cr+^{248}Cm}$ and $\mathrm{^{50}Ti+^{249}Cf}$ for E120. Neither E119 nor E120 has been observed yet. There were also several theoretical studies~\cite{Kayumov2022, Santhosh2017, Zhang2024} suggesting different reactions for the production of E119 and E120. The plans to conduct experiments on the synthesis of new elements were announced by JINR~\cite{ OGANESSIAN201562, Oganessian2009, oganessian2022, Oganessian2025}, LBNL~\cite{Gates2024}, and IMP (Lanzhou)~\cite{gan2022}.

Study of superheavy elements' chemical properties is also of great importance. The interest in studying properties of superheavy elements is largely driven by the desire to determine the limits of applicability of the periodic law, according to which  there is a smooth trend in properties of elements within the same group. However, it is not uncommon for some properties of superheavy elements to differ significantly from those of their lighter homologues. For example, in Refs.~\cite{eliav1996, kaygorodov2021,Guo:21}, it was shown that oganesson possesses a positive electron affinity, which fundamentally distinguishes it from the other noble gases. The difference in the electronic structure of superheavy elements from that of their lighter homologues is caused, on the one hand, by the growing influence of relativistic effects associated with increasing nuclear charge \cite{eliav2015}, and on the other hand, by the influence of electronic correlations associated with the increasing number of electrons. 

The number of papers studying properties of atomic E119 and molecules containing it were published previously. The ground state configuration of an atom of E119 was predicted to be [Rn]$5f^{14}6d^{10}7s^{2}7p^{6}8s^{1}$  (see, e.~g., Ref.~\cite{nefedov2006} and Refs. therein). Calculations of the ionization potential (IP), the electron affinity (EA), the removal energies, and the dipole polarizability of E119 were carried out in Refs.~\cite{tupitsyn2021, saetgaraev2024, landau2001, eliav2005EA, eliav2005IP, hangele2013, Dinh2008, Dzuba2013, Lim2005, borschevsky2013}. Сhemical properties of diatomic molecules E119H, E119F, and E119Cl were studied in Refs.~\cite{thierfelder2009, hangele2013, miranda2012, Pershina2025}. Adsorption properties of atomic E119, E119H, and E119OH on hydroxylated quartz and gold surfaces were studied in Refs.~\cite{Ilias2024, Pershina2025}. The properties of E119 and its compounds have been found to be determined mainly by the relativistic contraction of the $8s$-shell.

Calculations of the IP and EA of E119 have been performed in several previous studies~\cite{tupitsyn2021, saetgaraev2024, landau2001, eliav2005EA, eliav2005IP,  hangele2013, Dinh2008, Dzuba2013, Lim2005}. For correlation treatment, the framework of the configuration interaction method and many-body perturbation theory in the basis of the Dirac–Fock–Sturm orbitals (CI-DFS+MBPT) was applied in Ref.~\cite{tupitsyn2021}. Calculations of the IP and EA were carried out with the help of single-reference coupled cluster method in Refs.~\cite{saetgaraev2024, Lim2005}. Correlation potential method (CPM) was used to take into account the contribution of electron correlations in Refs.~\cite{Dinh2008, Dzuba2013}. Other studies~\cite{saetgaraev2024, landau2001, eliav2005EA, eliav2005IP,  hangele2013} employed the Fock-space coupled cluster method, in which only single and double excitations are typically included, because the treatment of higher‑order excitations is computationally demanding. As it was demonstrated in our recent work~\cite{saetgaraev2024}, the IP and EA of E119 calculated within the single-reference and multireference frameworks are in a good agreement. The remaining differences between the results of these methods can be primarily explained by the neglect of high‑order excitations. To obtain more accurate values for the IP and EA one has to include excitations higher than singles and doubles. In this study, we present calculations of the IP and EA of element 119 that include high‑order excitations, together with relativistic and quantum electrodynamic (QED) effects.

\section{Methods and computational details}\label{sec:methods}

As the initial approximation, we employ the Dirac–Coulomb (DC) Hamiltonian, given by

\begin{equation*}
\label{eq:HDC}
    H_{\text{DC}} = \Lambda^+ \left[\sum_i (c \vec{\alpha}_i \cdot \vec{p}_i + \beta_i c^2 + V_{\mathrm{nucl}}(i)) + \sum_{i < j} \frac{1}{ r_{ij}} \right] \Lambda^+,
\end{equation*}
where $ \vec{\alpha} $ and $ \beta $ are the standard four-dimensional Dirac matrices, $V_{\mathrm{nucl}}(i)$ is the nuclear potential (the finite nuclear size effect is taken into account using the Gaussian model of charge distribution \cite{VISSCHER1997}), the summation runs over all electrons in the system, and $ \Lambda^+ $ is the projector onto the positive-energy Dirac–Fock one-electron states.

To describe the electronic structure, we employ the relativistic single-reference coupled cluster (SR-CC) approach  accounting for single, double, and perturbative triple cluster amplitudes [SR-CCSD(T)]~\cite{Visscher:96a}. In addition, we use the relativistic Fock-space coupled cluster (FS-CC) method~\cite{Kaldor:91,Visscher:01,Eliav:Review:22} (and Refs. therein) with single and double excitations (FS-CCSD). In the latter model, the E119$^+$ charge state was chosen as the Fermi vacuum. Low-lying electronic states of E119 and E119$^-$ correspond to the $0h1p$ and $0h2p$ sectors of the Fock space, respectively. The active space comprised $8s$ spinor. Different SR-CC and FS-CC calculations were performed with varying numbers of correlated electrons, virtual-orbital energy cutoffs, and basis sets.
In the final calculations, 100, 101, and 102 electrons were correlated for the cation ([Rn]$5f^{14}6d^{10}7s^{2}7p^{6}$, $\mathrm{E119^{+}}$), the neutral atom ([Rn]$5f^{14}6d^{10}7s^{2}7p^{6}8s^{1}$, $\mathrm{E119}$), and the anion ([Rn]$5f^{14}6d^{10}7s^{2}7p^{6}8s^{2}$, $\mathrm{E119^{-}}$), respectively. 
Virtual spinors above 300 a.~u. were truncated. For the final calculations, we used a basis set obtained by extending the $41s36p25d22f4g4h2i$ set optimized in our previous study~\cite{saetgaraev2024}. The resulting uncontracted basis set consists of $41s36p25d22f9g6h5i$ Gaussian functions. Dirac–Hartree–Fock (DHF) and SR-CCSD(T) calculations as well as integral transformation were performed with the DIRAC code~\cite{DIRAC19, Saue2020}, while FS-CCSD calculations were carried out using the EXP-T program package~\cite{Oleynichenko:website, Oleynichenko:2020}.

To account for electron correlation beyond the SR-CCSD(T) model, we included contributions up to triple and perturbative quadruple cluster amplitudes in the SR-CCSDT(Q) method~\cite{Kallay:6}. The computational cost of this method drastically increases with the size of the basis set, making the constructing of compact basis sets and the assessment of their practical applicability of considerable importance. For these calculations, we used the generalized (Gatchina) relativistic pseudopotential (GRPP) method~\cite{Titov1999, Petrov2004, Mosyagin2016, Oleynichenko2023}, in which the GRPP for E119 was generated in Ref.~\cite{mosyagin2020} for 9 active electrons. The GRPP approach allows one to use contracted basis sets, in which each basis function is expressed as a linear combination of several primitive Gaussians, whereas in the DIRAC code~\cite{DIRAC19, Saue2020} employed in this study, only uncontracted basis sets composed of primitive Gaussian functions can be used for the case of four-component calculations of heavy elements. The contracted basis set was generated following the procedure described in Refs.~\cite{SKRIPNIKOV2013,Athanasakis_Kaklamanakis_2025,Oleynichenko2024} for constructing compact atomic natural orbital (ANO)-like sets. The construction began with obtaining relativistic one-particle reduced density matrices (1-RDMs) from two-component SR-CCSD calculations for $\mathrm{E119^{+}}$, $\mathrm{E119}$, and $\mathrm{E119^{-}}$ electronic charge states. The 1-RDMs in the atomic orbital (AO) representation were averaged over these electronic states, spin blocks, and orbital projections of the AO basis functions. The resulting effective density matrix was treated as the density-operator matrix in the AO representation, and its eigenvectors with the largest eigenvalues (occupation numbers) were used to construct the compact basis set employed in the SR-CCSDT(Q) calculations.

The {\sc dirac} program package~\cite{DIRAC19, Saue2020}, supplemented by the LIBGRPP library~\cite{Oleynichenko2023}, was used to perform the DHF and two-component GRPP calculations and to carry out the subsequent transformation of the molecular integrals. The EXP-T program~\cite{Oleynichenko:website, Oleynichenko:2020} was used to solve the SR-CCSD amplitude equations and to construct the density matrices. The ANO-type contracted basis set was generated with the {\sc natbas} code~\cite{SKRIPNIKOV2013, Athanasakis_Kaklamanakis_2025, Oleynichenko2024}. The high-order excitation correction was estimated as the difference between the SR-CCSDT(Q) and SR-CCSD(T) results obtained with the 9-electron GRPP method. These calculations were performed using the MRCC program~\cite{Kallay2001, Kallay2003}.

The leading two-electron correction to the DC Hamiltonian is the Gaunt term:
\begin{equation}
\label{Gaunt}
V_{ij}^{\mathrm{G}} = - \frac{(\bm{\alpha}_i \cdot \bm{\alpha}_j)}{r_{ij}}.
\end{equation}
This term represents the dominant contribution of the Breit electron–electron interaction. The Gaunt interaction was included during the construction of the one-electron basis set in the DHF procedure, followed by the relativistic mean-field eXact-2-Component (X2Cmmf) Hamiltonian transformation designed to reproduce accurately the positive-energy spectrum of the DHF equation~\cite{Sikkema2009, Kutzelnigg:2005, Ilias:2007}. The Gaunt electron-electron interaction contribution was evaluated at the FS-CCSD level as the difference between the total energies obtained with and without the Gaunt operator in the X2Cmmf Hamiltonian.

QED effects beyond the Dirac–Coulomb–Breit Hamiltonian can be approximately evaluated by means of the model-QED-operator~\cite{shabaev2013, shabaev2015}, local radiative potential method~\cite{Flambaum2005, Ginges2016PRA} and its implementation in the {\sc dirac} program~\cite{Sunaga2022}, nonlocal QED potential~\cite{tupitsyn2013}, local model potential~\cite{Pyykko2003, Pyykko2012} or the direct scaling of the Lamb shift results for H-like ions~\cite{Indelicato1990, Draganic2003, LOWE2013}. The QED operator consists of vacuum polarization (VP) and self-energy (SE) terms. To calculate the QED corrections, we used the model QED operator within the QEDMOD framework presented in Refs.~\cite{shabaev2013, shabaev2015} and an alternative implementation from Ref.~\cite{Skripnikov2021Ra+}. Calculations with the first implementation~\cite{shabaev2013, shabaev2015} were carried out using the configuration-interaction method in the basis of Dirac–Fock–Sturm orbitals (CI-DFS)~\cite{tupitsyn2003, tupitsyn2005, tupitsyn2018}. Calculations with the second implementation~\cite{Skripnikov2021Ra+, Skripnikov2021Ba+} were performed with correlation effects included at the all-electron SR-CCSD(T) and FS-CCSD levels. The QED contribution was obtained as the difference between the total energies calculated with and without the inclusion of these operators in the many-electron calculations.

\section{Results and discussion}

\subsection{One-electron basis set and high-order excitations}

As the first step of our computational scheme, we ana\-ly\-zed the influence of the choice of one-electron spinors used to construct Slater determinants on the results for high-order correlation effects. A small dependence of the final calculated parameters on the choice of spinor set can be considered as an additional test of the accuracy and reliability of the computational scheme~\cite{Skripnikov2015ThO}.

For this purpose, we performed SR-CCSD(T) and FS-CCSD calculations of the IP and EA using the DC Hamiltonian with the $41s36p25d22f2g2h$ basis set. This basis set was obtained by reducing the number of $g$, $h$, and $i$ functions in the optimized $41s36p25d22f4g4h2i$ basis set developed in Ref.~\cite{saetgaraev2024}. The SR-CCSD(T) calculations were carried out using two different schemes for constructing DHF orbitals. Within the first scheme, separate DHF calculations were performed for each charge state: $\mathrm{E119^{+}}$, $\mathrm{E119}$, and $\mathrm{E119^{-}}$. The second scheme implies the use of the same set of orbitals obtained from the DHF calculation for $\mathrm{E119^{+}}$ was used for all charge states. The results are presented in Table~\ref{Diff_clust_ampl}.

The values of the IP and EA computed by the SR-CCSD(T) and FS-CCSD methods using the same orbitals for all states differ by 0.017 eV and 0.005 eV, respectively. The IPs calculated using different orbital construction schemes within the SR-CCSD framework are in a good agreement; however, the EA values differ from each other by 0.015 eV. Inclusion of perturbative triple excitations increases the discrepancy for both the IP and EA. To address this issue, we consider higher-order excitations within the SR-CCSDT(Q) approach in the GRPP framework. For this purpose, a compact basis set consisting of $10s8p8d6f2g$ contracted functions was constructed. This basis was used to calculate the contributions to the IP and EA from iterative triple and non-iterative quadruple excitations for both schemes. These corrections were then added to the SR-CCSD(T) values presented in Table~\ref{Diff_clust_ampl}. The discrepancy in EAs computed with different schemes decreases significantly when full iterative triple cluster amplitudes are included, becoming as small as 0.003 eV. Perturbative quadruple cluster amplitudes further slightly reduce the remaining difference. For the IP, the inclusion of high-order excitations does not affect the already small difference between the two schemes.

\begin{table*}
\centering
\setlength{\tabcolsep}{8pt}
\caption{IP and EA of E119 calculated using different sets of DHF orbitals: Set 1 -- orbitals from three separate DHF calculations for $\mathrm{E119^{+}}$, $\mathrm{E119}$, and $\mathrm{E119^{-}}$; Set 2 -- orbitals from a single DHF calculation for $\mathrm{E119^{+}}$. All values are given in eV.}
\label{Diff_clust_ampl}
\begin{tabular}{lrrrr}
\hline
Property    & \multicolumn{2}{c}{IP} & \multicolumn{2}{c}{EA} \\
Orbitals    & Set 1      & Set 2     & Set 1      & Set 2     \\ \hline
FS-CCSD     &            & 4.7798    &            & 0.7075    \\
SR-CCSD     & 4.7349     & 4.7358    & 0.6017     & 0.5872    \\
SR-CCSD(T)  & 4.7939     & 4.7971    & 0.6754     & 0.7022    \\
SR-CCSDT $-$ SR-CCSD(T)    & $\mathbin{-}$0.0017    & $\mathbin{-}$0.0089   & $\mathbin{-}$0.0041    & $\mathbin{-}$0.0335   \\
SR-CCSDT(Q) $-$ SR-CCSDT        & +0.0031    & +0.0031   & +0.0069    & +0.0105   \\
SR-CCSDT(Q) & 4.7953     & 4.7913    & 0.6782     & 0.6792    \\ \hline
\end{tabular}
\end{table*}

The contributions to the IP and EA from non-iterative and iterative triple excitations, as well as perturbative quadruple excitations, are given in Table~\ref{Diff_clust_ampl}. The IP and EA obtained within the second scheme are more sensitive to higher-order excitations. Therefore, the first scheme was adopted for the final all-electron calculations. The uncertainty of the computed IP and EA associated with higher-order excitations was estimated from the perturbative quadruple-excitation corrections in Table~\ref{Diff_clust_ampl}, amounting to 0.0031 and 0.0069 eV, respectively. The IP and EA values obtained at the SR-CCSDT(Q) level for different sets of one-electron spinors agree within these uncertainties.

\subsection{Basis set and correlation convergences}

To address the convergence of results with respect to the basis set size and estimate the associated uncertainties, we performed calculations of the IP and EA using the DC Hamiltonian, the SR-CCSD(T) method, and three Gaussian basis sets: $41s36p25d22f2g2h$ (initial), $41s36p25d22f4g4h2i$, and $41s36p25d22f9g6h5i$ (final). The results are summarized in Table~\ref{Diff_basis_sets}. As additional functions are incorporated into the basis set, the IP increases, whereas the EA is less sensitive with respect to the further addition of basis functions. The EA decreases with the first augmentation ($+2g2h2i$) and then increases with the second ($+5g2h3i$).

\begin{table}[H]
\centering
\setlength{\tabcolsep}{8pt}
\caption{IP and EA of E119 calculated at the SR-CCSD(T) level with the initial $41s36p25d22f2g2h$ primitive Gaussian basis set, and contributions from additional Gaussian functions. All values are in eV.}
\label{Diff_basis_sets}
\begin{tabular}{lrr}
\hline
Basis set & \multicolumn{1}{c}{IP} & \multicolumn{1}{c}{EA} \\ \hline
Initial    & 4.7939                 & 0.6754                 \\
$+2g2h2i$ & +0.0006                & $\mathbin{-}$0.0012    \\
$+5g2h3i$ & +0.0044                & +0.0017                \\
Final     & 4.7989                 & 0.6759                 \\ \hline
\end{tabular}
\end{table}

We divided the uncertainty due to the basis set incompleteness into contributions from $spdf$-, $ghi$-, and higher-angular-momentum functions. The $spdf$ contribution was estimated as a difference between two scalar-relativistic pseudopotential calculations with the $21s20p17d15f4g4h2i$ and $24s24p24d23f4g4h2i$ primitive Gaussian basis sets. The $21s20p17d15f4g4h2i$ set was obtained by removing primitive Gaussians with exponential parameters $\alpha \geq 500$ from the $41s36p25d22f4g4h2i$ basis set, as they are unnecessary for the GRPP approach. The $24s24p24d23f4g4h2i$ set differs from $21s20p17d15f4g4h2i$ in that all primitive Gaussians with $\alpha \leq 10$ are replaced by an even-tempered set. The {\sc cfour} code \cite{cfour}  was used for these calculations. Variation of the $spdf$ functions yields $4\cdot10^{-6}$~eV for the IP and 0.0004~eV for the EA. The uncertainty due to $ghi$-function variation was estimated as the difference between the SR-CCSD(T) results with the $41s36p25d22f4g4h2i$ and $41s36p25d22f9g6h5i$ basis sets
with the DC Hamiltonian, amounting to 0.0044~eV (IP) and 0.0017~eV (EA). The higher-angular-momentum contribution was evaluated as the difference between the FS-CCSD results within the X2Cmmf Hamiltonian framework obtained with the $41s36p25d22f9g6h5i$ and $41s36p25d22f9g6h5i2k$ basis sets, giving 0.0006~eV and 0.0004~eV for the IP and EA, respectively. 
To estimate the error arising from freezing 18 innermost core electrons, we computed the IP and EA at the SR-CCSD(T) level with the DC Hamiltonian and the $41s36p25d22f4g4h2i$ basis set, correlating 101 and 119 electrons for the neutral atom. The difference between these calculations reaches 0.0004~eV for the IP and 0.0002~eV for the EA.

\subsection{Gaunt and QED corrections}

The Gaunt correction was computed at the FS-CCSD level with the X2Cmmf Hamiltonian. The QED correction was obtained using two different implementations of the model QED operator presented in Refs.~\cite{shabaev2013, shabaev2015} and in Refs.~\cite{Skripnikov2021Ra+, Skripnikov2021Ba+}. Calculations of the QED correction have been performed at the CI-DFS, SR-CCSD(T), and FS-CCSD levels. The results are presented in Table~\ref{QED}. We estimate the uncertainty of each QED correction calculation to be $\sim10$\%. Our results are consistent with each other within estimated uncertainties. The final values of IP and EA include the QED corrections obtained using the CI-DFS method: 0.0126(13)~eV for the IP and 0.0032(3)~eV for the EA. The contribution of the Breit interaction to IP was calculated in Refs.~\cite{eliav2005IP, Dinh2008, Dzuba2013} and found to be approximately 0.0030--0.0043~$\mathrm{eV}$, which is consistent with the value of the Gaunt contribution obtained in the present work (0.0038~$\mathrm{eV}$). QED effects for the IP and EA of E119 were also evaluated in Refs.~\cite{tupitsyn2021, eliav2005IP, eliav2005EA, hangele2013, Dinh2008, Dzuba2013, Ginges2016JPB, shabaev2013, Pykko1998, Labzowsky1999, Pyykko2003, Sapirstein2002, Ginges2016PRA}.
The QED correction to the IP obtained in the present study (0.0126~$\mathrm{eV}$) can be compared with the values reported in these studies, which range from 0.0083~$\mathrm{eV}$~\cite{Dinh2008} to 0.0188~$\mathrm{eV}~$\cite{Labzowsky1999}. The QED correction to the EA reported in Ref.~\cite{eliav2005EA} is 0.0123~$\mathrm{eV}$, compared to our value 0.0032~$\mathrm{eV}$.
\vspace{-0.5cm}
\begin{table}[H]
\centering
\setlength{\tabcolsep}{3pt}
\caption{Contributions to the IP and EA of E119 from Gaunt and QED effects calculated at the CI-DFS, SR-CCSD(T), and FS-CCSD levels. All values are in eV.}
\label{QED}
\begin{tabular}{llccc}
\hline
\multicolumn{1}{c}{Property} & \multicolumn{1}{c}{Contribution} & CI-DFS & SR-CCSD(T) & FS-CCSD \\ \hline
\multirow{2}{*}{IP}          & Gaunt                            &        &            & $\mathbin{-}$0.0038  \\
                             & QED                              & $\mathbin{-}$0.0126 & $\mathbin{-}$0.0113     & $\mathbin{-}$0.0111  \\ \hline
\multirow{2}{*}{EA}          & Gaunt                            &        &            & $\mathbin{-}$0.0005  \\
                             & QED                              & $\mathbin{-}$0.0032 & $\mathbin{-}$0.0030     & $\mathbin{-}$0.0036  \\ \hline
\end{tabular}
\end{table}
\vspace{-0.7 cm}
\subsection{Final values}
\vspace{-0.7 cm}

\begin{table}[H]
\centering
\setlength{\tabcolsep}{8pt}
\caption{Contributions to the total uncertainties of the final IP and EA values of E119. All values are in eV.}
\label{Error}
\begin{tabular}{lrr}
\hline
Error            & \multicolumn{1}{c}{IP} & \multicolumn{1}{c}{EA} \\ \hline
$spdf$-functions & <$10^{-4}$               & 0.0004                 \\
$ghi$-functions  & 0.0044                 & 0.0017                 \\
$k$-functions    & 0.0006                 & 0.0004                 \\
18 core electrons  & 0.0004                 & 0.0002                 \\
high-order excitations  & 0.0031                 & 0.0069                 \\
QED effects             & 0.0013                 & 0.0003                 \\ \hline
Total           & 0.0056                 & 0.0071                 \\ \hline
\end{tabular}
\end{table}

Table~\ref{Error} gives the uncertainty budget for the IP and EA. In Table~\ref{final_value} we present the individual contributions to the IP and EA. The final values for the IP and EA are 4.7839~eV and 0.6750~eV, respectively. 

\begin{table}[H]
\centering
\setlength{\tabcolsep}{8pt}
\caption{Final values of the IP and EA of E119 with their contributions. All values are in eV.}
\label{final_value}
\begin{tabular}{lrr}
\hline
Method & \multicolumn{1}{c}{IP} & \multicolumn{1}{c}{EA} \\ 
\hline
FS-CCSD   & 4.7965                 & 0.7176                 \\
SR-CCSD(T)   & 4.7989                 & 0.6759                 \\
SR-CCSDT(Q) $-$ SR-CCSD(T)  & +0.0014                & +0.0028                \\
QED         & $\mathbin{-}$0.0126    & $\mathbin{-}$0.0032    \\
Gaunt       & $\mathbin{-}$0.0038    & $\mathbin{-}$0.0005    \\ 
Total, this work  & 4.7839(56)                 & 0.6750(71)                 \\ 
\hline
\multicolumn{3}{c}{Previous studies:} \\
CI-DFS+MBPT \cite{tupitsyn2021} & 4.768(100)  & 0.674(100)           \\
SR-CCSD(T) \cite{saetgaraev2024}     &  4.779(10)               & 0.671(4)          \\
FS-CCSD \cite{saetgaraev2024}     & 4.769                & 0.707          \\
FS-CCSD \cite{landau2001}     &                 & 0.7171          \\
IHFS-CCSD \cite{eliav2005EA}    &                 & 0.64870 \\ 
FS-CCSD \cite{eliav2005IP}     & 4.7829 &                 \\
FS-CCSD \cite{hangele2013}     & 4.7838           & 0.4850          \\
CPM \cite{Dinh2008}     & 4.8170           &          \\ 
CPM \cite{Dzuba2013}     & 4.7779           &          \\ \hline
\end{tabular}
\end{table}

The final estimates of uncertainties were obtained as the square root of the sum of the squares of the individual uncertainties. The two dominant contributions arise from the $ghi$-function basis set and the high-cluster-amplitude correlation uncertainties.

The results for the IP and EA of E119 obtained in the present work and the corresponding results from Refs.~\cite{tupitsyn2021, landau2001, eliav2005IP, eliav2005EA, hangele2013, Dinh2008, Dzuba2013} are presented in Table~\ref{final_value}. The DCB Hamiltonian was employed in all these studies. As shown in Table~\ref{final_value}, our IP value is in a good agreement with the corresponding results reported in Refs.~\cite{tupitsyn2021, eliav2005IP, hangele2013, Dzuba2013}. Within the estimated error bars, the EA does not agree with the values given in Refs.~\cite{landau2001, eliav2005EA, hangele2013}. The primary source of this discrepancy is the inclusion of higher-order excitations in our calculations, which were not accounted for in the previous works. Minor differences between our results and those of Refs.~\cite{tupitsyn2021, landau2001, eliav2005IP, eliav2005EA, Dinh2008, Dzuba2013} can also be attributed to the use of different basis sets, different values of the QED contribution to the IP and EA, and the inclusion of the retardation part of the Breit interaction in earlier studies.

The results reported in Ref.~\cite{hangele2013} are especially noteworthy. The IP reported in Ref.~\cite{hangele2013} is in a good agreement with other studies. However, the EA presented in Ref.~\cite{hangele2013} differs from the results of other studies by approximately 0.2 eV, which is approximately $30-40 \%$ of the total EA. The studies~\cite{landau2001} and~\cite{hangele2013} employ the same basis set taken from Ref.~\cite{malli1993}, the same computational package~\cite{landau1999}, and the same virtual space. The noticeable difference between Refs.~\cite{landau2001} and \cite{hangele2013} is the number of correlated electrons (50 and 28, respectively), but this factor cannot explain the observed discrepancy between EA values. Therefore, it remains unclear to us why the value of the EA in Ref.~\cite{hangele2013} differs so significantly from the other results.

As it can be seen from Table~\ref{final_value}, the EA uncertainty obtained in the present work (0.0071~eV) is larger than the corresponding value from our previous study~\cite{saetgaraev2024} (0.004~eV), (but the agreement between the EA values is reasonable within the estimated uncertainties). In the earlier work, we did not account for high-order excitations and considered only the basis set contribution to the IP and EA uncertainties. It is noteworthy that the basis set uncertainties listed in Table~\ref{Error} are smaller than our earlier estimates.

\section{Conclusion}

In this work, the IP and EA of E119 were computed  using the composite method based on the relativistic single-reference coupled cluster theory with single, double and perturbative triple excitations, SR-CCSD(T). Corrections from iterative triple and non-iterative quadruple excitations to the IP and EA of E119 were evaluated for the first time; to evaluate them the special compact basis set was constructed. Inclusion of high-order cluster amplitudes reduces the difference between calculations employing two different schemes for constructing DHF orbitals, thus verifying a good convergence with respect to the inclusion of electron correlation effects. The QED corrections to the IP and EA of E119 were calculated using two different implementations of QED model operators within the CI-DFS, SR-CCSD(T), and FS-CCSD methods, and the results obtained with the different methods are in a good agreement. A detailed uncertainty analysis was conducted, considering contributions from the basis set incompleteness issues, correlation, and QED effects. The final recommended IP and EA values for E119 are 4.7839(56)~eV and 0.6750(71)~eV, respectively.

\section*{Acknowledgments}

This work was supported by the Russian Science Foundation (Grant No. 22-62-00004, https://rscf.ru/project/22-62-00004/). Computations were held on the basis of the HybriLIT heterogeneous computing platform (LIT, JINR)~\cite{Anikina2025}.

\newpage

\bibliographystyle{h-physrev.bst}
\bibliography{lit}


\end{document}